\documentclass[alpha-refs]{oup-contemporary}

\usepackage{graphicx}
\usepackage{siunitx}
\usepackage{amsmath}
\usepackage{amssymb}
\usepackage{bbding} 
\usepackage{subfigure}

\newcommand{\RNum}[1]{\uppercase\expandafter{\romannumeral #1\relax}} 

\journal{jcde}
\title{A review on geometric constraint solving}
\doi{xxxxxxx}
\jyear{202X}
\jvolume{X}
\jnumber{X}
\jstartpage{X}
\jendpage{X}
\jpubdate{Day Month Year}

\author[1,{\authfn{1}}]{Qiang Zou}
\author[1]{Zhihong Tang}
\author[2]{Hsi-Yung Feng}
\author[1]{Shuming Gao}
\author[3]{Chenchu Zhou}
\author[1]{Yusheng Liu}

\affil[1]{State Key Laboratory of CAD$\&$CG, Zhejiang University, Hangzhou, 310027, China }
\affil[2]{Department of Mechanical Engineering, The University of British Columbia, Vancouver, BC V6T 1Z4, Canada}
\affil[3]{Xi’an Aerospace Propulsion Institute, Xi’an, 710100, China}

\authnote{\authfn{1}Corresponding author. E-mail: \href{mailto: john.qiangzou@gmail.com}{john.qiangzou@gmail.com}}

\papercat{Review article}

\runningtitle{Geometric Constraint Solving Review}

\begin{document}
\begin{frontmatter}{a}{b}{c}
\maketitle
\begin{abstract}
This paper presents a comprehensive review of geometric constraint solving in parametric computer-aided design (CAD), with the major focus on its advances in the last 15 years. Geometric constraint solving can date back to the very first CAD prototype, Sketchpad, in the 1960s, but serious research studies were carried out only after parametric CAD was introduced in the late 1980s. In the following 30-year history of GCS research, two development stages may be identified: (1) the first 15 years (late 1980s - mid 2000s) were primarily devoted to geometric constraint decomposition for well-constrained systems or those with only structural constraint dependencies; and (2) the second 15 years (late 2000s - now) have seen research efforts shifted towards classification criteria and decomposition algorithms for general constraint systems (with and without non-structural constraint dependencies). Most existing reviews focused on the first 15 years. The problem researched in the second 15 years is, however, equally important, considering that a manually specified constraint system usually contains under- and over-constrained parts, and that such parts must be correctly detected and resolved before numerical solving can work. In this regard, this review paper covers both stages and will discusses what has already been made possible for handling general constraint systems, what developments can be expected in the near future, and which areas remain problematic.
\end{abstract}

\end{frontmatter}


\setcounter{section}{0}
\section{Introduction}

Computer-aided design (CAD) has been an indispensable tool in engineering modeling \parencite{liu2021memory}, analysis \parencite{cottrell2009isogeometric}, optimization \parencite{rao2019engineering}, manufacturing \parencite{zou2014iso} and maintenance \parencite{su2020accurate}. The notion of CAD may date back to the 1960s \parencite{coons1963outline, coons1960computer}. Nevertheless, it was not until the introduction of parametric CAD modelers in the late 1980s \parencite{shah1998designing} that CAD was extensively recognized by the user/engineer community. The main practical benefit of parametric CAD lies in geometry reuse, which allows users to attain a family of geometry variants through tuning parameters embedded in the model. Parametric CAD is not a single technique but a set of techniques, and common to all parametric techniques is associativity that allows local parametric changes to propagate automatically. The associativity is often implemented as a system of geometric constraints relating geometric entities in the model. By changing parameter values and then re-evaluating this system, automatic change propagation can be attained. This evaluation process is often referred to as geometric constraint solving.

A geometric constraint system (GCS) involves a set of geometric entities such point, line, plane, and cylinder, as well as a set of geometric constraints between these entities such as distance and angle. These constraints can usually be translated to algebraic equations whose variables are the coordinates of the participating geometric entities. Solving GCS refers to the process of attaining a solution (or solutions) of the GCS. Since the solving is understood in the context of geometry, a solution corresponds to a valid instantiation of the geometric entities such that all the constraints are satisfied.

\begin{figure}[t] 
\centering
\includegraphics[width=0.5\textwidth]{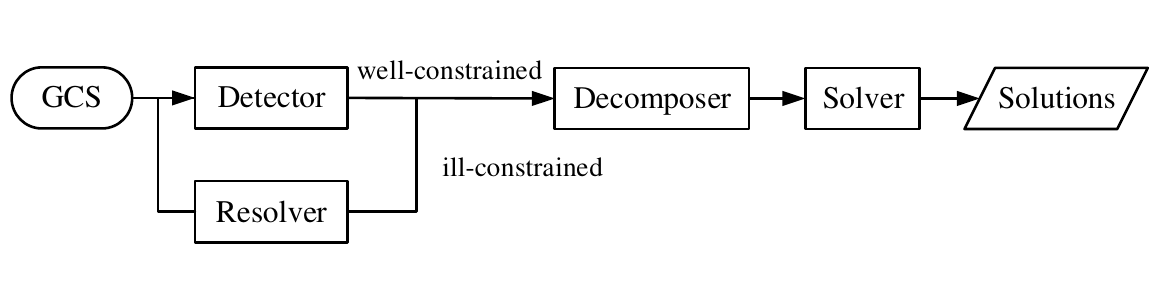}
\caption{Illustration of critical tasks in geometric constraint solving}
\label{fig:gcs-Process}
\end{figure}

Solving GCS is more than just applying numerical solvers such as the Newton's method. The GCS given through the user’s specification could contain under- and over-constrained parts \parencite{hoffmann2004Making,Zou2020adecision,hu2021geometric}. Such parts must be correctly detected and then resolved before numerical solvers can work, as shown in Fig.~\ref{fig:gcs-Process}. This gives rise to the need of generic criteria to characterize constraint states (under-, well-, and over-constrained), as well as robust detection and resolution algorithms to correct under-constrained and over-constrained parts in the given GCS. Another equally important issue is that directly solving GCS is computationally challenging. In view of this, many prefer decomposing a (well-constrained) GCS into small subsystems, whose solutions can then be assembled to attain the overall solution \parencite{bettig2011geometric}.

All in all, there are three fundamental problems in geometric constraint solving:
\begin{enumerate}
\item \textbf{GCS characterization:} effective criteria for characterizing under-, well-, and over-constraint states; 
\item \textbf{GCS detection:} extraction of ill-constrained parts in a given GCS. The essence is to extract the maximal well-constrained parts such that all under- and over-constrained parts emerge as isolated,  decoupled constraint groups \parencite{Zou2020adecision}; and
\item \textbf{GCS decomposition:} extraction of well-constrained parts in a given well-constrained GCS. Because the overall solving efficiency is governed by the largest subsystems, the essence here is to minimize the size of the well-constrained parts \parencite{hoffman2001decomposition1}.
\end{enumerate}

The above topics have been studied for decades, and there are several surveys made available, e.g., \cite{hoffman2001decomposition1,  hoffman2001decomposition2, hoffmann2005brief, jermann2006decomposition, bettig2011geometric}. Nevertheless, they primarily focused on decomposition methods for well-constrained GCS or those with only structural constraint dependencies, which was the research focus from the late 1980s through the mid 2000s. Since the mid 2000s, there have been research efforts devoted to effective/robust GCS classification criteria, detection, and decomposition algorithms for general constraint systems (with and without non-structural constraint dependencies). This work gives a comprehensive review on geometric constraint solving, covering both periods and for general GCS (with a hybrid state of under-, well-, and over-constraint). The current progresses and remaining issues will be emphasized, particularly for the recently developed witness configuration method (WCM) \parencite{michelucci2006geometric,thierry2011extensions}, a very promising method that can overcome limitations conventional methods suffer from.

The rest of this paper is organized as follows. Section 2 presents the fundamentals of GCS and GCS solving. Sections 3-6 presents several branches of solving approaches classified by different representation schemes of GCS; for each branch of approaches, the representation scheme, criteria for characterizing constraint states, detection, and decomposition are presented. Section 7 introduces the witness configuration method and presents the discussions on its limitations. The last section concludes this paper.

\section{Geometric constraint systems}

\subsection{GCS definitions}

In 2D and 3D parametric modeling, a geometric constraint system generally declares the geometric objects that must be included in the design model, and the geometric constraints that must be satisfied by these geometric objects. 

Specifically, the geometric objects are the basic geometric elements embedded at 2D or 3D Euclidean space, e.g. points, lines, and conics in 2D, and points, lines, planes, and surfaces in 3D. The geometric constraints describes the shape of geometric objects or the relationships among several geometric objects. The geometric constraint can be further classified into three categories: parametric constraints, non-parametric constraints, and algebraic constraints that relate the parameters from different parametric constraints. The parametric constraints (e.g. distance and angle) use parameters to define the dimensions of the model; the parameter values need assignments for creating the model, and need editing for modifying the shape of the model. The non-parametric constraints (e.g. perpendicular and parallel) define the structural relationships between two geometric objects; the non-parametric constraints can be regarded as the parametric constraints with parameters fixed at specific values. Algebraic constraints describe relationships among the parameters from different parametric constraints, e.g. $d_1=2d_2$ where $d_1$ and $d_2$ are the parameters from two distance constraints.


To define the shape of the model, the values of the parameters need to be assigned. The problem of geometric constraint solving is to find the solutions to the geometric constraint system, which are the coordinates of the geometric objects satisfying all the geometric constraints. Through examining the solution(s) of the GCS, notions of under-, well-, and over-constraint can be formally defined as follows, which is the base of the discussion in this work.

\textbf{Definition 1.}  Given a GCS and let $S$ be its solution space. The GCS is consistently constrained if $S\neq\emptyset$, and inconsistently constrained otherwise.

\textbf{Definition 2.}  Given a GCS and let $S$ be its solution space. The GCS is under-constrained if the cardinality $|S|$ is infinite.

\textbf{Definition 3.}  Given a GCS and let $S$ be its solution space. The GCS is consistently over-constrained if there exists a subset of the GCS such that the corresponding reduced equations have the same solution space as $S$.

\textbf{Definition 4.}  A GCS is over-constrained if it is inconsistently constrained or consistently over-constrained.

\textbf{Definition 5.}  A GCS is well-constrained if it is neither under-constrained nor over-constrained.

\begin{figure}[t] 
\centering
\includegraphics[width=0.5\textwidth]{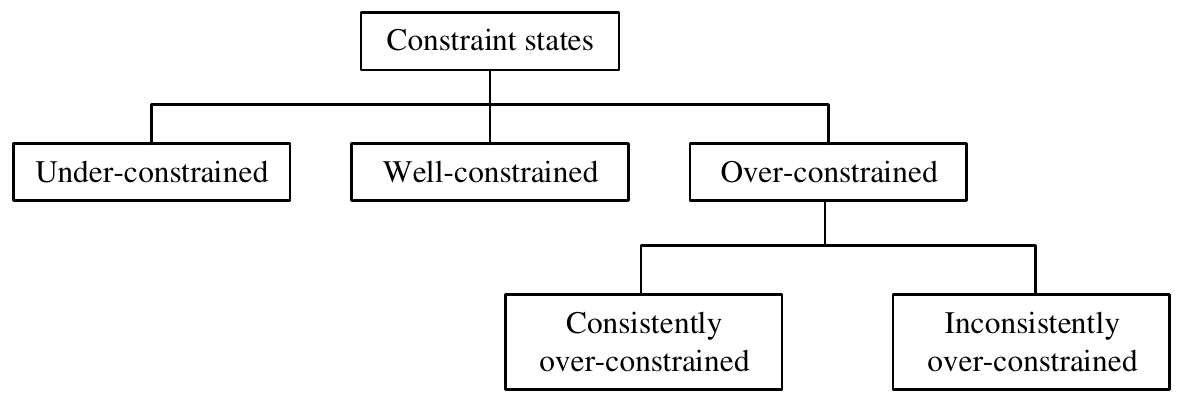}
\caption{The illustration of the constraint states}
\label{fig:2-States}
\end{figure}

The relationship among the listed constraint states are presented in ~Figure \ref{fig:2-States}. Generally, in a GCS problem, a well-constrained system or a consistently over-constrained system is preferred, since there are finite number of solutions to be found. For an under-constrained geometric constraint system, the infinite number of solutions cannot be obtained in a polynomial time; and for an inconsistently over-constrained geometric constraint system, no solution can be obtained, which makes the system meaningless. Therefore, it is a crucial step in the geometric constraint solving to determine the constraint state of the system, which is discussed in detail in Section 3.

\subsection{Solving process}

The process of geometric constraint solving is illustrated in ~Figure \ref{fig:gcs-Process}. Basically, the main components of the geometric constraint solver include: 

\begin{enumerate}
\item A detector. The detector determines whether the GCS in the current form is solvable or not.
\item A resolver. When the GCS is ill-constrained, the resolver provides guidance to the user for resolving the ill-constraints, or fix the GCS automatically.
\item A decomposer. For well-constrained GCS, the decomposer splits it into a set of subsystems for efficient solving.
\item A solver. It sequentially solves the subsystems, recombines them, and outputs the final solution(s) of the GCS.
\end{enumerate}

On this basis, the main steps of the geometric constraint solving are presented as follows:

\textbf{Step 1: sketching.} The user gives a sketch that declares the geometric objects and their topological relationships, and then adds geometric constraints.

\textbf{Step 2: constraint state characterization.} The detector determines the constraint state of the GCS. If the GCS is ill-conditioned (generally under-constrained or inconsistently over-constrained), go to Step 3; else, go to Step 4.

\textbf{Step 3: repairing.} The GCS is repaired automatically or by the user.

\textbf{Step 4: decomposition.} The GCS is decomposed into a set of subsystems. Also, a sequence of solving subsystems is generated, as well as a construction plan to recombine them.

\textbf{Step 5: subsystem solving.} The subsystems are solved according to the sequence, and the results are recombined as the final solution(s).

The critical issues of the geometric constraint solving are constraint state characterization and the decomposition. For this, literature on addressing these two issues are reviewed in Section 2 and Section 3.

\section{Algebraic approaches}

\subsection{Representation scheme}

The algebraic approaches basically translate the geometric constraints into a system of equations. The equations can be either linear or non-linear. It takes the coordinates of the geometric objects as variables; and its roots correspond to the solutions of the GCS. An example of the translation is given in ~Figure \ref{fig:2-1}(a). It shows a 2D geometric constraint system constructing a triangle. The geometric objects include 3 points ${P_1, P_2, P_3}$ and 2 lines ${L_1, L_2}$; the geometric constraints include 3 distances with the parameters ${d_1,d_2}$, an angle with the parameter ${\alpha}$, and 4 point-on-line constraints. Note that the line $\overline{P_1P_3}$ is not included for simplicity, since it can be directly defined once the positions of $P_1$ and $P_3$ are determined. The equation-based representation of the GCS in ~Figure \ref{fig:2-1}(a) is given as follows:

\begin{figure}[t] 
\includegraphics[width=0.45\textwidth]{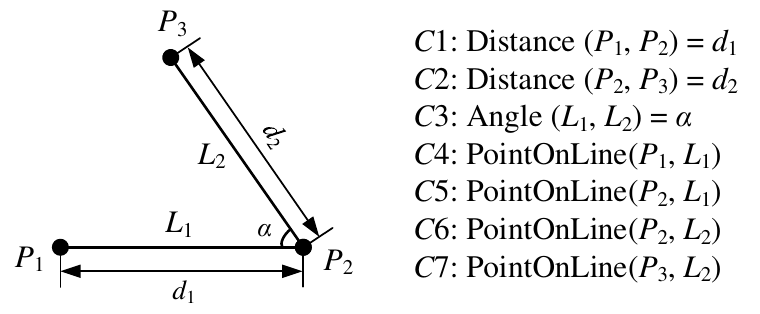}
\caption{A 2D geometric constraint system}
\label{fig:2-1}
\end{figure}

\begin{equation}
    \begin{cases}
       \text{$E_1$:  } (x_2-x_1)^2+(y_2-y_1)^2=d_1^2,\\
       \text{$E_2$:  } (x_3-x_2)^2+(y_3-y_2)^2=d_2^2,\\
       \text{$E_3$:  } (\varphi_1-\varphi_2)^2=(\pi-\alpha)^2,\\
       \text{$E_4$:  } x_1\cos{\varphi_1}-y_1\cos{\varphi_1}-\varrho_1=0,\\
       \text{$E_5$:  } x_2\cos{\varphi_1}-y_2\cos{\varphi_1}-\varrho_1=0,\\
       \text{$E_6$:  } x_2\cos{\varphi_2}-y_2\cos{\varphi_2}-\varrho_2=0,\\
       \text{$E_7$:  } x_3\cos{\varphi_2}-y_3\cos{\varphi_2}-\varrho_2=0,\\
    \end{cases}
    \label{eq:2.1}
\end{equation}
where the geometric objects and their coordinates are defined as $P_1(x_1,y_1)$, $P_2(x_2,y_2)$, $P_3(x_3,y_3)$, $L_1(\varphi_1,\rho_1)$, $L_2(\varphi_2,\rho_2)$. The values of the parameters ${d_1, d_2, \alpha}$ need to be assigned before solving, e.g. $d_1=10$, $d_2=10$, and $\alpha=\frac{\pi}{4}$. Moreover, additional equations need to be added to fix the Cartesian coordinate system with the geometric constraint system. For example, the following fixing equations can be added:

\begin{equation}
    \begin{cases}
       x_1=0,\\
       y_1=0,\\
       y_2-y_1=0,
    \end{cases}
    \label{eq:2.2}
\end{equation}
where the first two equations set $P_1$ as the origin, and the third equation takes the vector from $P_1$ to $P_2$ as the direction of $X$-axis. 

Given a system of equations with parameters assigned, algebraic solving approaches need to be employed to attain the root (or roots). Generally, the approaches can be classified into two branches, numerical approaches and symbolic approaches. 

\subsection{Numerical approaches}

The numerical approaches solve equation systems in a iterative manner, and yield approximations of the solutions in numerical schemes. The numerical approaches are practical in computation.

\subsubsection{Constraint state characterization}

The numerical approaches determine the constraint states of GCS by analyzing Jacobian matrix of the equation system. Generally, the Jacobian matrix expresses the variation of the geometries' coordinates under a slight change of constraint parameters (assume that the parameters are at the right side of equations). The Jacobian matrix, denoted as $J$, is an ($m\times n$) matrix that contains partial derivatives of each constraint equation:

\[
J = \begin{bmatrix}

    f_{11} & f_{12} & \dots & f_{1n} \\

    f_{21} & f_{22} & \dots & f_{2n} \\
    
    \vdots & \vdots & \ddots & \vdots \\

    f_{m1} & f_{m2} & \dots & f_{mn} \\

\end{bmatrix}_{m \times n}
\]
where
$f_{ij}=\frac{\partial F}{\partial x_{ij}}$.

A necessary and sufficient condition of a well-constrained GCS proposed by Light and Gossard \parencite{light1982modification} is that the the Jacobian has full rank at the solutions. On this basis, the constraint state criterion is carried out as follows:

\textbf{Criterion} Given the equation system of a GCS:

(1) If the number of equations is less than the number of variables, the GCS is under-constrained;

(2) If the Jacobian at the solutions is non-singular, and the number of equations is equal to the number of variables, the GCS is well-constrained; 

(3) If the Jacobian at the solutions is non-singular, and the GCS is not under-constrained, the GCS is over-constrained.

However, since the solutions are unknown beforehand, the criterion can hardly be applied directly. For this, several alternatives are proposed, which are introduced in the next subsection.

\subsubsection{Constraint state detection}

An alternative is to valuate and analyze Jacobian at random configurations instead, but singular configurations may be encountered where rank deficiency happens to the Jacobian.

\begin{figure}[t] 
\centering
\subfigure[]{\includegraphics[width=0.2\textwidth]{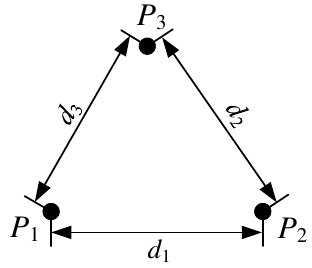}}
\centering
\subfigure[]{\includegraphics[width=0.2\textwidth]{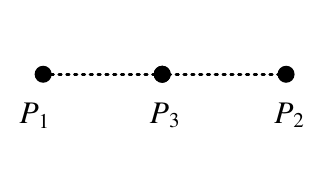}}
\caption{A 2D geometric constraint system and a configuration with singular Jacobian: (a) the geometric constraint system, and (b) the configuration with singular Jacobian.}
\label{fig:3-singular}
\end{figure}

For example, ~Figure \ref{fig:3-singular}(a) presents a geometric constraint system with 3 points $P_1(x_1,y_1)$, $P_2(x_2,y_2)$, $P_3(x_3,y_3)$ and 3 distances $d_1=10$, $d_2=10$, $d_3=10$ between the points. The equation system can be obtained as follows:

\begin{equation}
    \begin{cases}
        x_1=0,\\
        y_1=0,\\
        y_2-y_1=0,\\
        (x_2-x_1)^2+(y_2-y_1)^2=10^2,\\
        (x_3-x_2)^2+(y_3-y_2)^2=10^2,\\
        (x_1-x_3)^2+(y_1-y_3)^2=10^2.
    \end{cases}
    \label{eq:3-singular}
\end{equation}

The Jacobian is presented in ~Table \ref{tab:Jacobian}. A root of the equation system is given as $P_1(0,0)$, $P_2(20,0)$, $P_3(10,0)$. The Jacobian evaluated at the root is given in ~Table \ref{tab:nonsingular} with full rank; However, when evaluating the Jacobian at the configuration in ~Figure \ref{fig:3-singular}(b), the result becomes singular, presented in ~Table \ref{tab:singular}. 

\begin{table*}[bt!]
\caption{Jacobian of $F(X)$ in ~Equation (\ref{eq:3-singular})}
\label{tab:Jacobian}
\begin{tabular}{ccccccc}
\hline
& $\dot{x_1}$ & $\dot{y_1}$ & $\dot{x_2}$ & $\dot{y_2}$ & $\dot{x_3}$ & $\dot{y_3}$\\ 
\hline
$F_1$ & 1 & 0 & 0 & 0 & 0 & 0 \\ 
$F_2$ & 0 & 1 & 0 & 0 & 0 & 0 \\ 
$F_3$ & 0 & -1 & 0 & 1 & 0 & 0 \\ 
$F_4$ & $-2(x_2-x_1)$ & $-2(y_2-y_1)$ & $2(x_2-x_1)$ & $2(y_2-y_1)$ & 0 & 0 \\ 
$F_5$ & 0 & 0 & $-2(x_3-x_2)$ & $-2(y_3-y_2)$ & $2(x_3-x_2)$ & $2(y_3-y_2)$ \\ 
$F_6$ & $2(x_1-x_3)$ & $2(y_1-y_3)$ & 0 & 0 & $-2(x_1-x_3)$ & $-2(y_1-y_3)$ \\ 
\hline
\end{tabular}
\end{table*}

\begin{table}[bt!]
\caption{Non-singular Jacobian at the solution in ~Figure \ref{fig:3-singular}(a)}
\label{tab:nonsingular}
\begin{tabular}{ccccccc}
\hline
& $\dot{x_1}$ & $\dot{y_1}$ & $\dot{x_2}$ & $\dot{y_2}$ & $\dot{x_3}$ & $\dot{y_3}$\\ 
\hline
$F_1$ & 1 & 0 & 0 & 0 & 0 & 0 \\ 
$F_2$ & 0 & 1 & 0 & 0 & 0 & 0 \\ 
$F_3$ & 0 & -1 & 0 & 1 & 0 & 0 \\ 
$F_4$ & -20 & 0 & 20 & 0 & 0 & 0 \\ 
$F_5$ & 0 & 0 & 10 & -17.32 & -10 & 17.32 \\ 
$F_6$ & -10 & -17.32 & 0 & 0 & 10 & 17.32 \\ 
\hline
\end{tabular}
\end{table}

\begin{table}[bt!]
\caption{Singular Jacobian at the configuration in ~Figure \ref{fig:3-singular}(b)}
\label{tab:singular}
\begin{tabular}{ccccccc}
\hline
& $\dot{x_1}$ & $\dot{y_1}$ & $\dot{x_2}$ & $\dot{y_2}$ & $\dot{x_3}$ & $\dot{y_3}$\\ 
\hline
$F_1$ & 1 & 0 & 0 & 0 & 0 & 0 \\ 
$F_2$ & 0 & 1 & 0 & 0 & 0 & 0 \\ 
$F_3$ & 0 & -1 & 0 & 1 & 0 & 0 \\ 
$F_4$ & $-20$ & 0 & $20$ & 0 & 0 & 0 \\ 
$F_5$ & 0 & 0 & $-20$ & 0 & $20$ & 0 \\ 
$F_6$ & $-20$ & 0 & 0 & 0 & $20$ & 0 \\ 
\hline
\end{tabular}
\end{table}

Therefore, the difficulty lies in finding the non-singular configurations where the Jacobian has the same structures as that at solutions. To achieve this, several methods has been proposed.

\textbf{1. Sampling-based method }

The basic idea of sampling-based method is to increase the probability of finding non-singular configurations by increasing the number of sampling configurations. Since most of configurations are non-singular except for some certain points within the variable space, this method performs well practically. If a non-singular sampling configuration is find, then the GCS is determined as well-constrained. 

Haug et al. \parencite{haug1989computer} presented a perturbation method to deal with singular configurations. It obtains more configurations for valuation of Jacobian matrix.

Foufou et al. \parencite{foufou2005numerical} extended the numerical probabilistic method (NPM) which valuates the Jacobian matrix at a set of random configurations in the variable space.
Such methods are useful piratically, but theoretically they cannot guarantee to avoid singular configurations.

\textbf{2. Witness configuration method}

Instead of selecting configurations randomly, Michelucci et al. \parencite{michelucci2006geometric, michelucci2009interrogating} suggested to valuate the Jacobian at the witness configurations that has the same structure of Jacobian as that of solutions. To guarantee the same structure of Jacobian between witness configurations and solutions, all the singular constraints, e.g. point/line/plane incidences, needs to be satisfied in witness configurations. The aim is to avoid degenerate cases between geometric entities which may change the Jacobian structure. 

When all the singular constraints are stated by the user, a witness configuration can be computed as a solution to a GCS containing only these singular constraints. This GCS is generally highly under-constrained, and can be solved easily.

Li et al. \parencite{li2001numerical} presented a method to detect numerical redundancies. It gives a perturbation to the parameter value of each constraint, and then tests the solvability of the new GCS by optimization method \parencite{ge1999geometric}. This method can only handle consistent over-constrained GCS.

\subsubsection{Solving}

The earliest proposed methods and prototypes \parencite{sutherland1964sketchpad, Hillyard1978Characterizing, borning1981programming} employed the relaxation method, which starts with a initial guess of the solution, and iteratively gives perturbations to the variable values until some measure is minimized. The convergence of the relaxation method is generally slow. 

Later on, Newton-Raphson iteration was widely used in the subsequent work \parencite{light1982modification, lin1981variational, nelson1985juno}. It converges faster but heavily relies on an appropriate initial guess. It can hardly handle consistently over-constrained GCSs, since a square Jacobian is required within the iteration process.  Generally, the relaxation method and Newton-Raphson iteration can only find one root. 

For exhaustive searching of multiple solutions, homotopy continuation was applied \parencite{Allgower1993homotopy, lamure1996solving, durand2000systematic, imbach2017homotopy}. It theoretically guarantees to find all solutions to an equation system. However, as there is always a trade-off between efficiency and exhaustive searching, its computational load is heavier. 

Nearly 2000, numerical optimization was introduced to geometric constraint solving \parencite{ge1999geometric, Cao2006evolutionary, Yuan2006TPSO}. It is stable and less sensitive to the initial guess. The main advantage is that consistently over-constrained and under-constrained can be effectively handled. 

The numerical approaches are general, but they are time-consuming without involving constraint decomposition, especially for large-scale systems. Actually, the numerical approaches are commonly included in other decomposition approaches to solve subproblems, or are chosen as the last choice when other approaches do not work well.

\subsection{Symbolic approaches}

The symbolic approaches employ symbolic techniques to convert an equation system into a triangular form. The conversion, in some sense, is a decomposition of the GCS, where the equations in the triangular form can be solved sequentially. Representative methods include Gr\"{o}bner bases \parencite{bose1995grobner, buchanan1993constraint, kondo1992algebraic} and Wu-Ritt decomposition \parencite{chou1988introduction, chou1988mechanical}. 

\subsubsection{Constraint state characterization}

Basically, the symbolic approaches characterize the constraint states of the GCS by the triangular form of equations. 

\textbf{Criterion} Given a GCS $S$, let $P_1$ and $P_2$ be two points in $S$. Fixing the position of $P_1$ and the direction $P_1P_2$, $S$ is well-constrained iff the triangular set contains ($2n-4$) polynomials; $S$ is under-constrained iff the triangular set contains less than ($2n-4$) polynomials; $S$ is consistently over-constrained iff the triangular set contains more than ($2n-4$) polynomials; $S$ is inconsistently over-constrained iff there is no triangular set. 

\subsubsection{Constraint state detection and solving}

In symbolic approaches, the solving process is generally mingled with the constraint state detection process. Gr{\"o}bner basis method \parencite{buchberger1985Grobner, bose1995grobner} and Wu-Ritt method \parencite{wu2012mechanical, chou1988introduction, gao1998solving} are respectively introduced as follows.

Gr{\"o}bner basis method computes the Gr{\"o}bner basis for a system of equations.  The criterion of determining the constraint state of a GCS is presented as follows:

\textbf{Criterion} Assume a set of polynomials $f_0,f_1,\dots,f_s\in \mathbb{C}[x_1,x_2,\dots,x_n]$. The reduced Gr{\"o}bner basis $rgb_0$ of the ideal $<f_1,f_2,\dots,f_s>$ satisfies $rgb_0\neq {1}$ and $rgb_0\neq {0}$ with respect to any ordering. The new reduced Gr{\"o}bner basis of the ideal $<f_0,f_1,f_2,\dots,f_s>$ is $rgb_{new}$. If $rgb_{new}={1}$, the new added $f_0=0$ is a conflicting equation; if $rgb_{new}\equiv{0}$, $f_0=0$ is a redundant equation, $rgb_{0}\subset rgb_{new}$, $f_0=0$ is an independent equation.

The algorithm of generating Gr{\"o}bner basis is first proposed by \parencite{buchberger1985Grobner}. Kondo \parencite{kondo1992algebraic} introduced Gr{\"o}bner basis into geometric constraint solving and constraint management. Solving GCS problem using Wu-Ritt method is firstly introduced by Gao and Chou \parencite{gao1998solving}. Given a set of polynomials $P$, it decomposes a system of polynomial equations into a triangular set, and The zero set of the polynomials can be denoted as:

\begin{equation}
    Zero(P)=\cup_{1\leq i\leq k} Zero(TS_{i}/I[i]),
\end{equation}
where $TS_{i}$ is a polynomial set in triangular form, $I[i]$ is the production of initials of the polynomials in $TS_{i}$, and $k$ denotes the zero sets.The symbolic approaches are capable of finding all roots, but are rarely employed in practical GCS problems due to their high computational load.

\section{Geometric approach}

\subsection{Representation scheme}

The geometric approaches analyze and solve GCS in a geometrical way. These approaches directly operate with the original representation scheme of GCS, i.e. the full geometric descriptions of the geometric objects and the geometric constraints. Faced with these, the supporting theories and techniques include elementary geometry and 2D ruler-and-compass drawing. For example, the GCS in ~Figure \ref{fig:2-1} can be trivially solved geometrically by ruler-and-compass drawing, and the construction steps are given as follows:

\begin{enumerate}
    \item $\text{FixOrigin}(P_1)$;
    \item $\text{FixX-Axis}(L_1)$;
    \item $P_2=\text{CreatePoint}(P_1,L_1,d_1)$;
    \item $L_2=\text{CreateLine}(P_2,L_2,\alpha)$;
    \item $P_3=\text{CreatePoint}(P_2,L_2,d_2)$.
\end{enumerate}

However, the supporting theories and techniques could have little effects when handling more complicated situations, e.g. 3D geometric constraint solving and complex 2D problems. 

\subsection{Constraint state characterization}

The strategy of the geometric approaches is to decompose the whole GCS into a set of subsystems that can be trivially analyzed and solved. Generally, the decomposition process is executed first, and the detection is then operated on the decomposition results. The constraint states of GCS is characterized by the unmatched parts as follows:

\textbf{Criterion} Given a GCS and a repertoire of well-constrained patterns or construction rules to be matched with, when the matching terminates:

\begin{enumerate}
\item the system is well-constrained if all its components are matched;
\item the system is under-constrained if there are geometric objects left;
\item the system is over-constrained if there are geometric constraints left.
\end{enumerate}

\subsection{Constraint state deterction}

\subsection{Decomposition algorithms}

Basically, the constraint state of a geometric constraint system is determined by iteratively matching it with the repertoire of predefined well-constrained patterns or construction rules. These well-constrained patterns, as well as the construction rules, are the abstractions of the existing geometric knowledge. The process terminates when no patterns or rules are applicable. 

The rule-based method employs geometric assertions and axioms to identify rigid patterns in the geometric constraint system. 

Sunde \parencite{sunde1987cad} proposed a method to handle 2D geometric constraint problems with points constrained by distances and angles. First, basic patterns in the geometric constraint system are created, e.g. a pair of points with a distance constraint (called a CD) and two pair of points with an angle constraint (called a CA). Then, several rules are used to iteratively merge the existing patterns, e.g. three CDs pairwise sharing one point or two CAs sharing one segment can be combined into a bigger rigid subsystem. The method terminates when all the points belong to a CD set. Essentially, the rules are abstracted from the geometric knowledge of triangles. The subsequent literatures further extended the types of geometric entities and geometric constraints, and the repertoire of rigid patterns  \parencite{verroust1992rule,joan1997correct,Joan1997arulerandcompass}. Particularly, Joan-Arinyo and Soto-Riera \parencite{joan1997correct,Joan1997arulerandcompass} introduced the point-line distance constraint (called a CH). 

Another set of methods \parencite{aldefeld1988variation,bruderlin1989rule,bruderlin1990symbolic,sohrt1991interaction,yamaguchi1990constraint} develop an axiomatic description of elementary geometry for ruler-and-compass construction. These axioms are expressed as rewrite-rules for replacing constraints which are stored in a database by equivalent ones. Essentially, these methods realize the automatic process of ruler-and-compass drawing. As the rule-and-compass drawing generates the geometric loci of elements, an important issue is to handle the loci intersection problem. For this, Gao et al. \parencite{gao2004solving} proposed a method to calculate the locus intersection in 2D by discretization. 

The decomposition process is essentially an automatic geometric reasoning process, thus it directly works at the original representation scheme. The reliability of the methods mainly depends on the completeness of the repertoire, i.e. whether all well-constrained patterns can be enumerated, or whether there are a set of basic rules to construct all kinds of well-constrained systems.

\section{Topological approach}

\subsection{Representation scheme}

The other set of approaches choose to translate the geometric constraint system into a graph, so that graph theories and algorithms are available to further analyze the structural properties and decompose the system. Specifically, the graph-based representation of the geometric constraint system can be further classified into two categories: the equation graph and the constraint graph.

\begin{figure}[t] 
\centering
\subfigure[]{\includegraphics[width=0.48\textwidth]{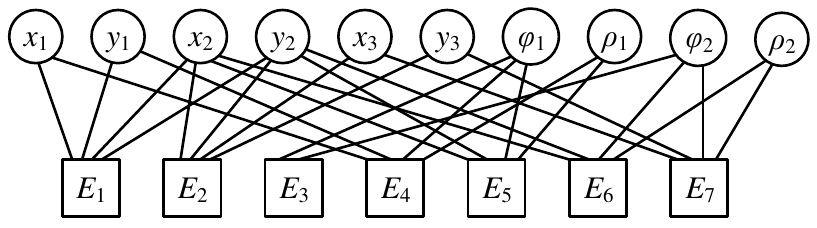}}\\
\subfigure[]{\includegraphics[width=0.35\textwidth]{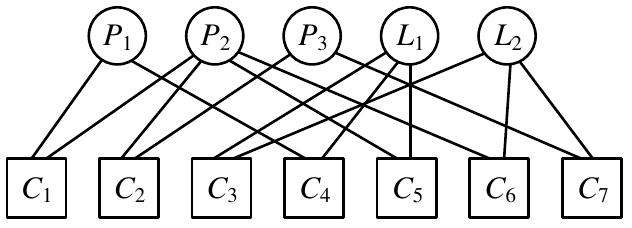}}\\
\subfigure[]{\includegraphics[width=0.2\textwidth]{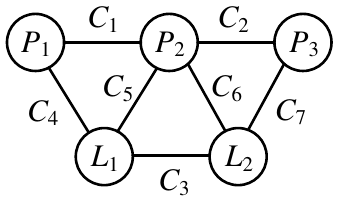}}
\caption{The graph representations of the geometric constraint system in ~Figure \ref{fig:2-1}(a): (a) the equation graph, (b) the constraint graph where constraints are defined as nodes, and (c) the constraint graph where the constraints are defined as edges.}
\label{fig:2-2}
\end{figure}

The equation graph is derived from the equation-based representation. A node denotes either a variable or an equation; and an edge linking a variable node and an equation node denotes that the variable is constrained by the equation. Since the nodes can be divided into two groups linked by the edges, the equation graph can also be represented as a bipartite graph. For example, the equation graph derived from ~Equation \ref{eq:2.1} is presented in ~Figure \ref{fig:2-2}(a).

The constraint graph is directly derived from the original scheme of the geometric constraint system. It abstracts geometric objects by their degree of freedom (DOF), and abstracts geometric constraints by thier degree of constraints (DOC). The DOF of a geometric object is the number of independent coordinates used to define it; and the DOC of a geometric constraint is the number of equations needed to defined it. For example, a 2D point and a 2D line both have 2 DOFs, and a 2D circle has 3 DOFs; a distance constraint with non-zero parameter value has 1 DOC. Generally, the constraint graph has two representation schemes. One is the general constraint graph where the nodes only represent geometric objects and the edges are the constraints; and the other is the bipartite graph where a node denotes either a geometric object and a geometric constraint, and an edge links a geometric constraint and a geometric object it constraints. ~Figure \ref{fig:2-2}(b-c) respectively presents the two schemes of the constraint graph.

\subsection{Constraint state characterization}

\subsubsection{Criteria for constraint graph}

Laman's theorem \parencite{laman1970graphs} characterizes the rigidity of 2D bar frameworks, i.e. geometric constraint systems composed of points and distance constraints between the points.

\textbf{Theorem 1.} A 2D GCS composed of $n$ points constrained by $m$ distances is rigid iff $2\times n-m=3$, and for any subsystem composed of $n'$ points and $m'$ distances, $2\times n'-m'\geq 3$.

However, the theorem is restricted in 2D points and distance constraints, thus cannot be applied in more complicated systems. To extend the theorem, Podgorelec et al. \parencite{Podgorelec2008dealing} proposed the concept of structural rigidity in the geometric constraint problems.

\textbf{Definition (Structural constraint states)} Let $S=(C,X,A)$ be a constraint system. 

1) The system $S$ is \textit{structurally over-constrained} iff there exist a subsystem $S'=(C',X',A')$ of $S$ such that $DOF(X')-DOC(C')<D$.
       
2) The system $S$ is \textit{structurally well-constrained} iff it is not structurally over-constrained and $DOF(X)-DOC(C)=D$.
    
3) The system $S$ is \textit{structurally under-constrained} iff it is not structurally over-constrained and $DOF(X)-DOC(C)>D$.

\textbf{Definition (Structurally over-constrained)} A constraint $e$ is a structural over-constraint if a structurally over-constrained subsystem $G'=(V',E')$ of $G$ with $e\in E'$, can be derived such that $G"=(V',E'-e)$ is structurally well-constrained.

\begin{figure}[t] 
\centering
\subfigure[]{\includegraphics[width=0.2\textwidth]{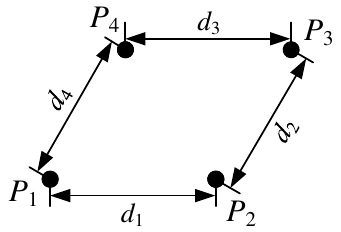}}
\subfigure[]{\includegraphics[width=0.2\textwidth]{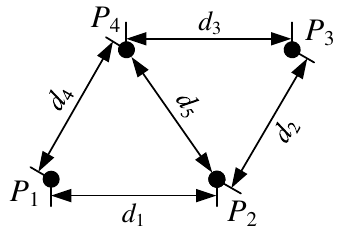}}
\subfigure[]{\includegraphics[width=0.2\textwidth]{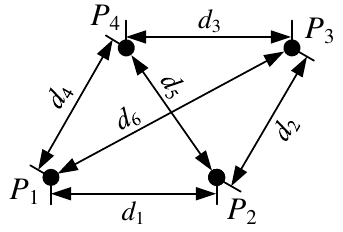}}
\caption{Examples of structural constraint states: (a) a structurally over-constrained system, (b) a structurally well-constrained system, and (c) a structurally under-constrained system.}
\label{fig:5-structuralconstraint}
\end{figure}

Examples are presented in ~Figure \ref{fig:5-structuralconstraint} to illustrate the definition. The 2D geometric constraint system in the figure is composed of 4 points. Since each point has 2 DOFs, the whole system has 8 DOFs totally. As shown in ~Figure \ref{fig:5-structuralconstraint}(a), the system is structurally under-constrained. 4 distance constraints are set; as each constraint gives 1 DOC, the whole system satisfies $2\times 4-1\times 4>3$. As shown in ~Figure \ref{fig:5-structuralconstraint}(b), the system is structurally well-constrained, because the whole system satisfies $2\times 4-1\times 5=3$, and the two subgraphs $P_1P_2P_4$ and $P_2P_3P_4$ both satisfy $2\times 3-1\times 3=3$. As shown in ~Figure \ref{fig:5-structuralconstraint}(c), the system is structurally over-constrained, because the whole system satisfies $2\times 4-1\times 6<3$. 

\begin{figure}[t] 
\centering
\subfigure[]{\includegraphics[width=0.2\textwidth]{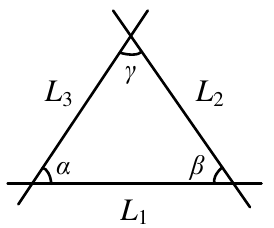}}
\subfigure[]{\includegraphics[width=0.15\textwidth]{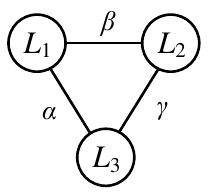}}
\caption{A structurally well-constrained but geometrically over-constrained geometric constraint system: (a) the geometric constraint system, and (b) the constraint graph.}
\label{fig:5-structgeoinconsis}
\end{figure}

Note that a structural well-constrained system can be geometrically over-constrained. An example is presented in ~Figure \ref{fig:5-structgeoinconsis}. The geometric constraint system in ~Figure \ref{fig:5-structgeoinconsis}(a) is composed of 3 lines $L_1$, $L_2$, and $L_3$ constrained by 3 angles $\alpha$, $\beta$, and $\gamma$; the constraint graph illustrated in ~Figure \ref{fig:5-structgeoinconsis}(b) shows that the system is structurally well-constrained. However, there is a implicit constraint $\alpha+\beta+\gamma=180^{\circ}$, making the system geometrically over-constrained. 

\begin{figure}[t] 
\centering
\subfigure[]{\includegraphics[width=0.2\textwidth]{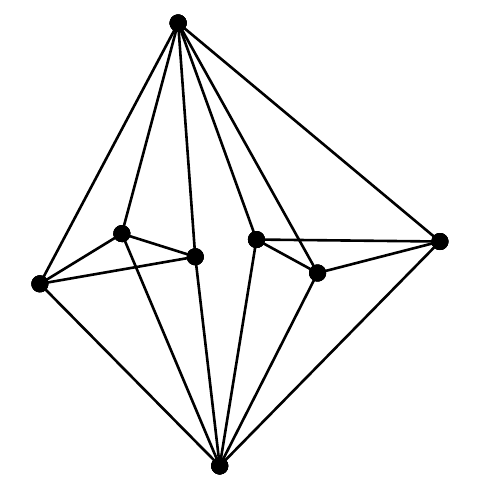}}
\subfigure[]{\includegraphics[width=0.15\textwidth]{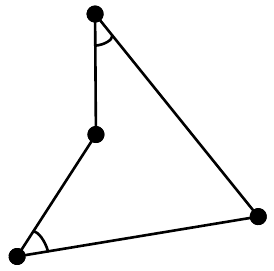}}
\subfigure[]{\includegraphics[width=0.15\textwidth]{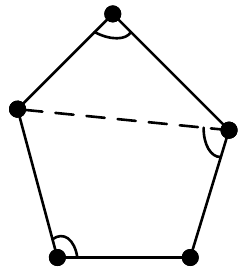}}
\subfigure[]{\includegraphics[width=0.18\textwidth]{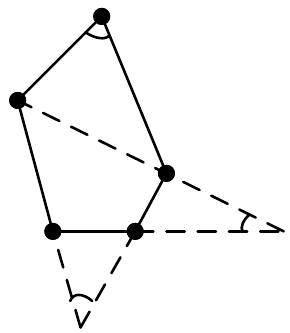}}
\caption{Counter examples of 3D extension of Laman's theorem: (a) Double-Banana, and (b-d) Ortuzar's configurations.}
\label{fig:5-CounterExamples}
\end{figure}

As for 3D extension of Laman's theorem, the term $2v-3$ should be replaced with $3v-6$, since each 3D point has 3 coordinates and a displacement of a 3D body is characterized by 6 parameters, i.e. 3 translation parameters and 3 rotation parameters. However, the condition is necessary but not sufficient. Counter examples are Double-Banana and Ortuzar's configurations, presented in ~Figure \ref{fig:5-CounterExamples}. The system is structurally well-constrained by our definition but is actually under-constrained. So far, no combinatorial characterizations have been found yet. 

Sitharam and Zhou \parencite{Sitharam2004atractable} proposed an approximate characterization of 3D rigidity, which effectively handles some geometry (e.g. Double-Banana) misjudged by Laman's theorem. 

\textbf{Definition (Structural constraint states)} Let $S=(C,X,A)$ be a constraint system. 

1) The system $S$ is \textit{structurally over-constrained} iff there exist a subsystem $S=(C,X,A)$ of $S$ such that $DOF(X)-DOC(C)<D$.
       
2) The system $S$ is \textit{structurally well-constrained} iff $DOF(X)-DOC(C)=D$ and all subgraphs satisfy $DOF(X)-DOC(C)\geq D$.
    
3) The system $S$ is \textit{structurally under-constrained} iff $DOF(X)-DOC(C)>D$.

Another problem arises that whether all rigid bodies need $D$ ($D=3$ in 2D and $D=6$ in 3D) parameters for displacements. Counter examples are 2 points binding with a distance constraint in 3D, or a 3D cylindrical surface. The former needs only 5 independent displacements since the body cannot rotate around the axis crossing the 2 points; the latter also needs only 5 independent displacements as it cannot rotate around its axis. That is to say, the number of independent displacements are determined by the system itself rather than the dimensions. To address this, degree of rigidity (DOR) \parencite{Jermann2003Algorithms} is proposed to replace the original $D$. The updated definitions are as follows:

\textbf{Definition (Structural constraint states)} Let $S=(C,X,A)$ be a constraint system. 

1) The system $S$ is \textit{structurally over-constrained} iff there exist a subsystem $S=(C,X,A)$ of $S$ such that $DOF(X)-DOC(C)<DOR(X)$.
       
2) The system $S$ is \textit{structurally well-constrained} iff $DOF(X)-DOC(C)=DOR(X)$ and all subgraphs satisfy $DOF(X)-DOC(C)\geq DOR(X)$.
    
3) The system $S$ is \textit{structurally under-constrained} iff $DOF(X)-DOC(C)>DOR(X)$.

For example, the DOR of 3 collinear 2D points is 2, but the DOR of 3 non-collinear 2D points is 3. This indicates that the value of DOR depends on the relative positions between the geometric objects in the rigid bodies. However, the relative positions cannot be directly obtained from the pure graph representation.

The main limitation of the criteria mentioned above is that they are restricted in specific geometries (e.g. point, lines, planes) and constraints (e.g. distances). Thus, extensions on types of geometries and constraints are needed for applications in practical CAD problems. 

\subsubsection{Criteria for equation graph}

For an equation graph, a variable has only one DOF, and an equation has only one DOC. The criterion of constraint states determination for an equation graph is presented as follows:

\textbf{Criterion} If there are only unsaturated constraints, the GCS is over-constrained; if there are only unsaturated variables, the GCS is under-constrained; if all variables and constraints are saturated, the GCS is well-constrained.

As the criterion works on the equation level, it is not limited to any specific types of geometries. However, it can hardly be applied to GCS with so called black box constraints, which cannot be converted to equations. 

\subsection{Constraint state detection}

For a equation graph, it is also converted into a bipartite graph, where maximal matching algorithms can be used. Such method is pioneered by Serrano \parencite{serrano1987constraint}. In \parencite{ait2014reduction}, the Dulmage \& Mendelsohn (D\& M) algorithm was used to divide the equation graph into a well-constrained part, an under-constrained part, and an over-constrained part. A maximal matching of an equation graph can be further decomposed into well-constrained subsystems by decomposing it into strongly connected components. An effective approach is a linear depth first search algorithm by Tarjan \parencite{Tarjan1972DepthFirstSA}.

For a constraint graph, it is first converted into a bipartite graph. Then, the problem is translated into a maximal flow problem where a number of methods is applicable. Such translation was first discussed by Hoffman et al. \parencite{hoffmann1997finding}, and was extended in \parencite{hoffman2001decomposition2}. The applied maximal flow methods are mainly based on greedy algorithms.

\subsection{Decomposition}

\subsubsection{Matching-based decomposition}

The matching-based approach employs matching algorithms to decompose the geometric constraint system. It directly acts on the graph representation. 

At first, the matching-based approach operates on the equation graph representation. This was pioneered by Serrano \parencite{serrano1987constraint}, which consists of the following steps:

(1) Find a maximum matching of the equation graph. 

(2) Transfer the equation graph into a digraph. Specifically, orient each edge in the maximum matching from the variable node to the equation node; and orient each unmatched edge from the equation node to the variable node.

(3) Identify strong connected components from the digraph and collapse them into super nodes.

(4) Obtain the solution plan by topological sorting.

This work was followed by Ait-Aoudia et al. \parencite{ait2014reduction}. Dulmage-Mendelsohn (D-M) decomposition was introduced that divided the whole equation graph into three sub-parts: the well-constrained sub-part, the over-constrained sub-part, and the under-constrained sub-part. With D-M decomposition, the ill-conditions of the geometric constraint system is made explicit to the user or the CAD system and specific treatment can be provided for repairing. Moreover, the maximum matching of the well-constrained sub-part is guaranteed to be a perfect matching.

Further, the matching-based approach was extended to constraint graph. Lamure and Middleditch \parencite{lamure1996solving} proposed a weighted maximal matching to the system decomposition. The method no longer works at equation level, but directly works on the constraint graph which is weighted by DOFs of geometric entities and DOCs of geometric constraints. The method converts the constraint graph into a directed one by maximal weighting. The directions of edges are set by the following rules: a edge is set bidirectional if its weight is non-zero, and is directed from geometric entity to constraint if its weight is zero. 

Hoffmann et al. \parencite{hoffmann1997finding} introduced the flow-based algorithm. The method first computes the maximum flow in the constraint graph and identifies the minimal dense subsystem. The minimal dense subsystem is then extended iteratively to include the neighboring nodes. Two extension algorithms are proposed, which are called condensing algorithm (CA) \parencite{hoffman2001decomposition1} and the frontier algorithm (FA) \parencite{hoffman2001decomposition2}. Both the algorithms collapse the subsystem, but differ in the schemes. Specifically, the CA simply collapse the identified subsystem into a super node, while the FA preserves the frontier vertices of the subsystem. Both the algorithms yield a cluster tree, which indicates a solving sequence. Moreover, work \parencite{hoffmann2004Making} has been proposed to develop the ability of FA in handling over-constrained systems.

\subsubsection{Bottom-up decomposition}

\begin{figure}[t] 
\centering
\subfigure[]{\includegraphics[width=0.2\textwidth]{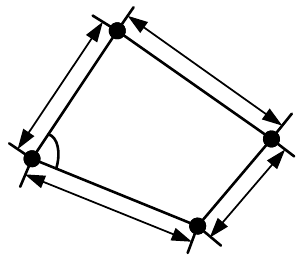}}
\subfigure[]{\includegraphics[width=0.35\textwidth]{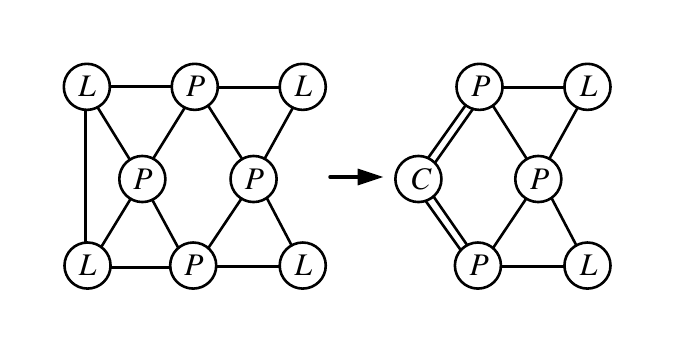}}
\caption{An example of bottom-up decomposition.}
\label{fig:bottom-up}
\end{figure}

The bottom-up approach iteratively aggregates components into bigger ones. It was pioneered by Bouma et al. \parencite{bouma1995geometric}. In this method, triangles in the constraint graph are identified and combined. An example of the decomposition process is illustrated in ~Figure \ref{fig:bottom-up}. Compared with top-down approach, the bottom-up approach behaves better in handling over-constrained problems. 

One of the critical issues to be addressed is extending the repertoire of rigid patterns. Subsequent work include handling variable-radius circles \parencite{owen1996constraints,hoffmann2002variable,chiang2004revisiting}, Bezier curves \parencite{Hoffmann1995geometricconstraints}, conics \parencite{Fudos1996conics}. Hoffmann and Joan-Arinyo \parencite{hoffmann1997symbolic} proposed a method to handle the geometric constraint systems involving algebraic constraints that relates parameters (e.g. two distance constraints $d_1$ and $d_2$ satisfy $d_1=2d_2$).

Bouma et al \parencite{bouma1995geometric} proposed a DOF-based method to make the bottom-up decomposition. The method operates on the constraint graph. The rigid subsystem is called a cluster. The main rule used in the method is "three clusters pairwise sharing an geometric element can be merged into a bigger cluster". However, this rule sometimes yield non-rigid subsystems, i.e. the subsystem formed by three lines with three angle constraints are not rigid.

Apparently, triangle is only one of the rigid patterns, and the main difficulty of the rule-based approach lies in a complete repertoire of patterns. Gao and Zhang \parencite{gao2003classification} presented a classification of basic merge patterns between two rigid bodies in 2D. For all the eleven cases presented, the symbolic solution of each case is given, and whether it can be constructed by ruler-and-compass method is analyzed.

Later on, non-rigid patterns are also taken into consideration. Shreck et al. \parencite{schreck2006geometrical,schreck2006using} extended the types of 2D patterns with invariant properties, e.g. similarity. van der Meiden and Bronsvoort \parencite{van2010non} considered 3D geometric constraint solving, and presented two non-rigid patterns: scalable clusters and redial clusters. Also, a mapping method was presented from 3D geometric constraint system with several kinds of primitives to the one with only 3D points.

Considered that not all geometric constraint systems are ruler-and-compass constructible \parencite{gao2003classification}, Mathis et al. \parencite{mathis2012decomposition} and Barki et al. \parencite{barki2016re} proposed a re-parameterization method to make the ruler-and-compass technique available. It is achieved by removing a constraint and adding a new constraint with driving parameter. With the driving parameter changing, a geometric locus is generated, from which the positions that satisfying the removed constraint are taken as solutions. The method essentially converts the well-constrained geometric constraint system into the 1-DOF one. The limitation is that only one removable constraint is considered at current stage; the situation becomes more complex when more than one constraint are removed.

\begin{figure}[t] 
\centering
\subfigure[]{\includegraphics[width=0.2\textwidth]{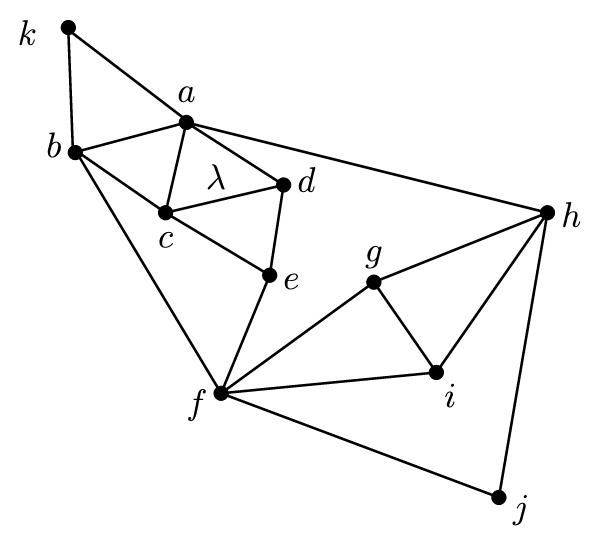}}\\
\subfigure[]{\includegraphics[width=0.4\textwidth]{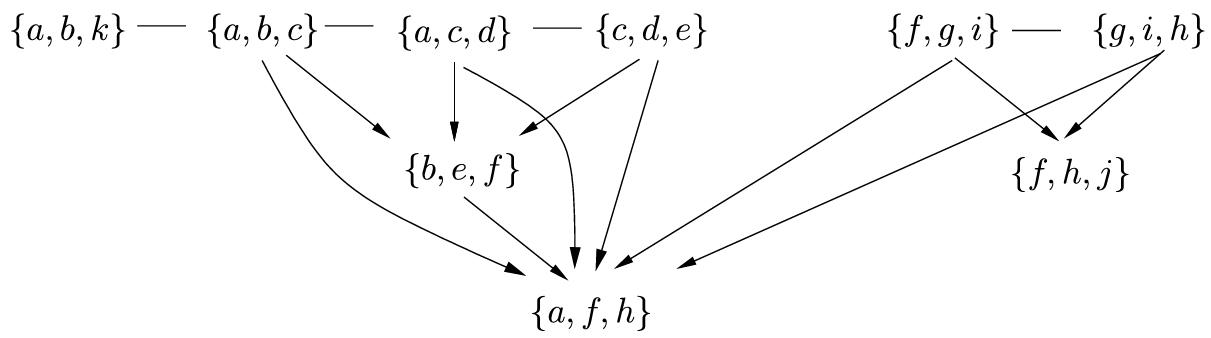}}
\caption{An example of h-graph: (a) The GCS, and (b) the corresponding h-graph.}
\label{fig:h-graph}
\end{figure}

The decomposition generates a ternary tree where each node captures a construction of a pattern (e.g. a triangle) and the leaves capture the set of geometric objects. The tree representation shows the construction sequence of the GCS. However, the representation doesn't explicitly show the dependencies between the construction steps. For this, Hidalgo and Joan-Arinyo \parencite{Hidalgo2015hgraph} introduced h-graph, which is obtained from the tree representation and captures the dependency relationships between construction steps. An example is given in ~Figure \ref{fig:h-graph}. In ~Figure \ref{fig:h-graph}(b), the undirected edges represent co-dependencies, and the directed edges represent indirect dependencies.

The bottom-up decomposition has a better performance in handling over-constrained GCSs, but can hardly handle under-constrained problems.

\subsubsection{Top-down decomposition}

Top-down decomposition methods work by iteratively splitting the whole geometric constraint system into smaller components. Apparently, the critical issue to be addressed lies in finding the splitting position.

\begin{figure}[t] 
\centering
\includegraphics[width=0.2\textwidth]{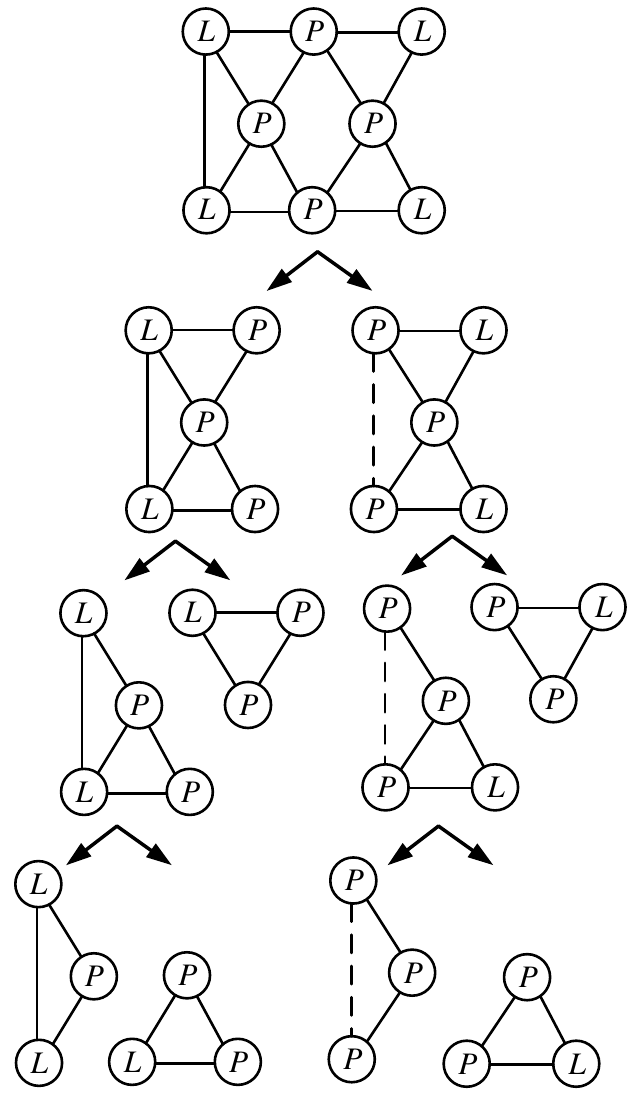}
\caption{An example of top-down decomposition.}
\label{fig:top-down}
\end{figure}

The top-down decomposition was pioneered by Owen \parencite{owen1991algebraic}. The method handles 2D geometric constraint systems with points and lines constrained by distance and angle constraints. During each iteration, this method decomposes the constraint graph at an articulation pair. The articulation pair is a pair of nodes that their removal results in two disconnected subgraphs. After splitting, the pair of nodes is duplicated in both subgraphs, and a virtual bond, i.e. a edge, is added between them. The virtual bonds constraints the relative position between the two vertices. The splitting terminates when no more articulation pairs can be found. After the splitting, the geometric constraint system is decomposed into triangles. The triangles can be solved according to the reverse order of the decomposition. An example is illustrated in ~Figure\ref{fig:top-down}.

Joan-Arinyo et al. \parencite{joan2004revisiting} presented s-tree to formalize the top-down decomposition. Fudos and Hoffmann \parencite{fudos1997graph} presented a similar decomposition method that combined bottom-up techniques to handle over- and under-constrained problems. Zhang and Gao \parencite{zhang2004geometric} proposed a method to decompose a GCS into a d-tree in a top-down manner. It extends the connection types to 3D. 

Generally, the top-down decomposition can better decompose the under-constrained GCS, because the under-constrained GCS only has more articulation pairs. However, the top-down decomposition can hardly handle over-constrained GCSs.

\subsubsection{Propagation-based decomposition}

The propagation-based approach aims at converting the constraint graph into a directional one where the incident edges of each vertex are all incoming edges except for one. More specifically, the approach includes DOF propagation and variable propagation.

Freeman-Benson et al. \parencite{freeman1990Incremental} proposed the DeltaBlue algorithm to incrementally solve the GCS by local propagation. Sannella et al. \parencite{sannella1993skyblue} presented the SkyBlue algorithm as a successor to the DeltaBlue algorithm. It can detect loops, but cannot entirely break them. Veltkamp et al. \parencite{Veltkamp1995Quantum} proposed a method to break loops. Borning et al. \parencite{borning1996indigo} proposed a local propagation method to deal with GCSs with inequalities.

The main drawback of this approach is that loops are often generated during the propagation. Veltkamp el al. \parencite{Veltkamp1995Quantum} proposed a method to break loops. Borning et al. \parencite{borning1996indigo} proposed a local propagation method to handle geometric constraint system with inequality constraints.

\section{Introduction and limitations of VCM}

\subsection{A brief introduction}
The above review suggests that conventional methods have inherent limitations and are often used beyond their validity. By contrast, the very recent WCM is likely the most promising method to bypass those limitations. Successful applications of this method have been reported, including  mechanism design, freeform surface design, \parencite{mathis2014coordinate,zou2019push,hu2017over}, to name a few.

VCM is essentially a bunch of criteria to characterize constraint states. It attains these criteria by examining how the constraint equations behave under infinitesimal perturbations made to the equation variables. That is, VCM applies a small perturbation $\Delta X$ to $X$ and examines the response $\Delta F=F(X+\Delta X)-F(X)$ of the constraint equations, which are related by (the first order approximation):

\begin{equation}
    \Delta F=J(X)\cdot\Delta X+O(\|\Delta X\|^2 ),
\end{equation}
where $J(X)$ is the Jacobian matrix evaluated at point $X$. The main result of VCM is that if $J$ is evaluated at a carefully selected point called witness configuration, the constraint dependencies (i.e., over-constraint) between the equations are described by the linearly dependent rows of the Jacobian matrix, i.e., the vectors in the kernel (or null space) of $J^T$. Under-constraint is closely related to the free perturbations that do not violate existing constraints, i.e., $\Delta F=J\cdot\Delta X=0$; that is, under-constraint is described by the vectors in the kernel of $J$. A configuration is deemed a witness configuration if certain types of constraints in the given model has already been satisfied by the configuration under perturbation, i.e., $X$, see \parencite{kubicki2014witness} for generation of such configurations. A detailed discussion of witness configurations and the necessity of them for VCM to work well can be found in \parencite{thierry2011extensions,  michelucci2009interrogating}.

The formal criteria of under-, well- and over-constraint given by VCM are as follows \parencite{thierry2011extensions}. A model is over-constrained if $Kernel(J^T)\neq\emptyset$, where $Kernel(\cdot)$ denotes the kernel of a matrix: $Kernel(A)={x|Ax=0}$. This criterion can be rewritten as:

\begin{equation}
    Rank(J)-RowSize(J)<0.
\end{equation}

A model is under-constrained if $Dim(Kernel(J))>6$, where $Dim(\cdot)$ is to get the dimension of a linear space and the number 6 represents the 6 rigid-body transformations. This criterion can be rewritten as:

\begin{equation}
    Dim(Kernel(J))=ColumnSize(J)-Rank(J)>6.
\end{equation}
Using the rank-nullity theorem \parencite{Strang2003Introduction}. Following immediately, the criterion for a well-constrained GCS is given as:

\begin{equation}
     \begin{cases}
        (Rank(J)-RowSize(J)=0,\\
        ColumnSize(J)-Rank(J)=6.
     \end{cases}
\end{equation}

\begin{figure}[bt!] 
\centering
\includegraphics[width=0.5\textwidth]{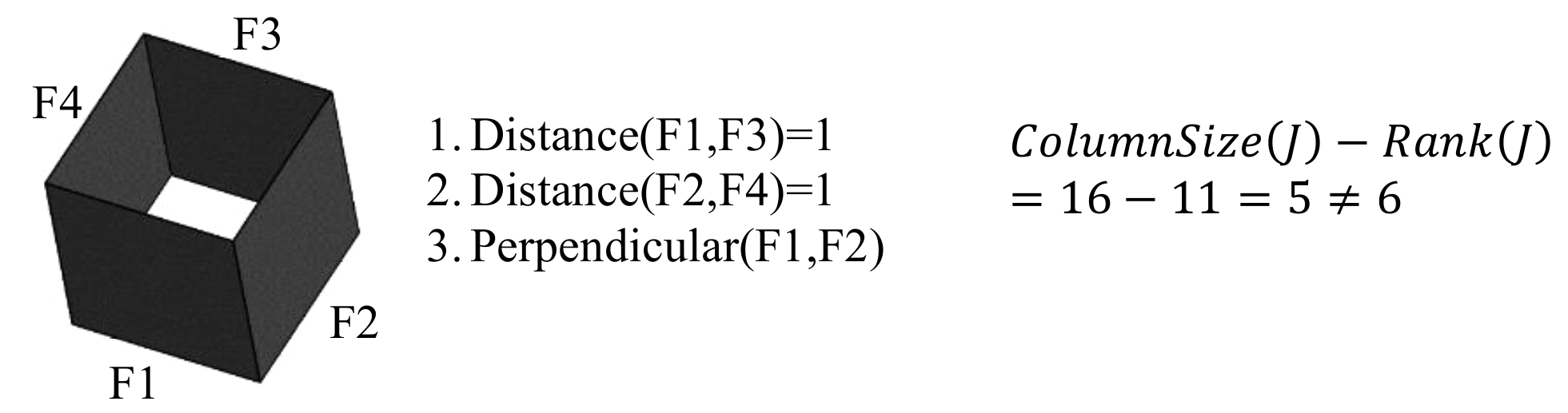}
\caption{A failure example: a well-constrained model but incorrectly characterized by VCM.}
\label{fig:vcm_1}
\end{figure}

These criteria were accepted once, but were later proven wrong. The failure is due to the assumption that a well-constrained GCS admits exactly 6 DOFs, which is seen to be not true in many real cases \parencite{Jermann2002anewstructural}. ~Figure \ref{fig:vcm_1} shows such a failure example\footnote{In this example, a plane is described with the tuple $(a,b,c,d)$ and the additional constraint $a^2+b^2+c^2=1$ (the plane equation is $ax+by+cz+d=0$). The equations of the involved constraints are as follows: (1) collinearity of two normal or direction vectors $v_1$, $v_2$ is represented with $v_1+tv_2=0$ where $t$ is a scalar unknown; and (2) point-to-point distance and vector angle are represented with dot products.}. To address this issue, the recent VCM replace the fixed number 6 with a variable called degree of rigidity (DOR) \parencite{michelucci2009interrogating}. The DOR notion may be first presented in \parencite{Jermann2002anewstructural}, and then was directly applied to VCM in \parencite{michelucci2009interrogating}. This notion essentially translates the 6 rigid-body transformations (3 axial translations and 3 axial rotations) to changes made to the parameters of the participating geometric entities, where these changes are stored as 6 vectors; then DOR of a GCS is given by the number of independent vectors among the 6 vectors, see Section 3 of \parencite{michelucci2009interrogating} for details. Consider, again, the example in ~Figure \ref{fig:vcm_1}. Its DOR is 5; then the numbers $ColumnSize(J)-Rank(J)$ and DOR are matched. This means that the constraint state of this example is characterized correctly with the help of the DOR notion.

\subsection{Limitation 0}

The most important limitation of the witness configuration method is that the subsystem constituting the witness configuration may not be well-constrained, and thus cannot be satisfied \parencite{zou2019limitations}.

\subsection{Limitation \RNum{1}}

The first limitation to be discussed in this subsection concerns with the ineffectiveness of VCM’s criteria for characterizing constraint states. Although enhanced with the DOR notion, there is still no rigorous proof for VCM. The introduction of the DOR notion is more like a patch for fixing the bugs encountered in the application of VCM than a systematic treatment of VCM’s limitations. In fact, none of the key publications on VCM, say \parencite{michelucci2006geometric, thierry2011extensions,  michelucci2009interrogating}, come with guarantees or with clearly stated limitations. Effectiveness in terms of constraint state characterization was often over-claimed in previous publications but was negated in later publications, which is the case for the introduction of the DOR notion to VCM.

\begin{figure}[bt!] 
\centering
\includegraphics[width=0.3\textwidth]{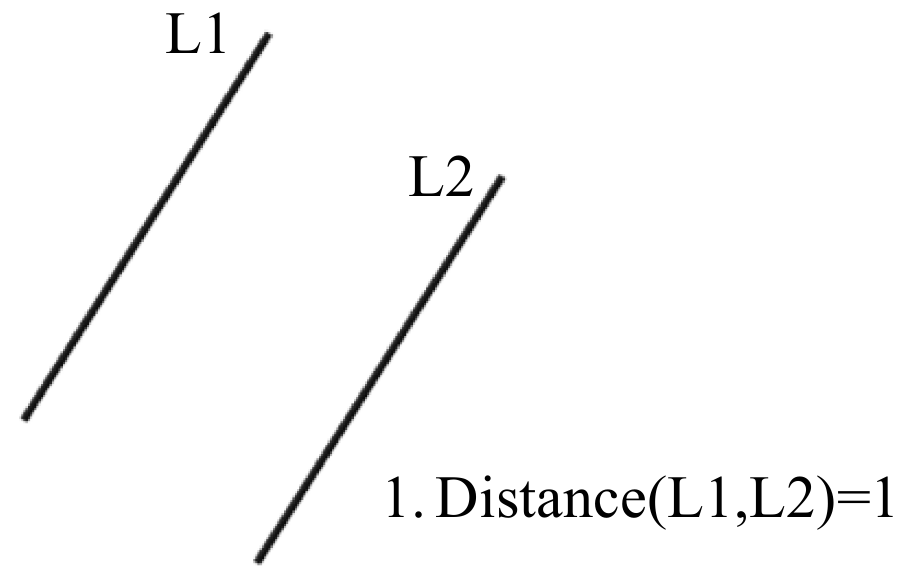}
\caption{Another failure example: a well-constrained model but incorrectly characterized by VCM.}
\label{fig:vcm_2}
\end{figure}

One limitation encountered by the authors in applying VCM is that the representation scheme used to describe the participating geometric entities has a significant effect on VCM’s effectiveness. Consider, for example, the four-plane configuration in ~Figure \ref{fig:vcm_1}, and change the original representation scheme to the following one: a plane is described with a point $p\in R^3$ on the plane and the normal $n\in S^2$ of the plane. The expression $ColumnSize(J)-Rank(J)$ becomes $24-13=11$, and DOR of this example becomes 6, which leads to a mismatch and a failed constraint state characterization. From this example, we can conclude that different representation schemes could result in different numbers of the terms $ColumnSize(J)$, $Rank(J)$, and DOR, and consequently could lead to opposite characterization results for a same GCS. ~Figure \ref{fig:vcm_2} gives one more example about such a situation, and ~Table \ref{tab:vcm_1} summaries the characterization results for this example and the one in ~Figure \ref{fig:vcm_1}. Representation \#1 refers to the original representation scheme; Representation \#2 refers to the scheme we just mentioned above; and the representation scheme used to describe the lines of the example in ~Figure \ref{fig:vcm_2} is: a line is represented by a point $p\in R^3$ on the line and the direction $d\in S^2$ of the line. Based on the results in ~Table \ref{tab:vcm_1}, it is safe to say that the current VCM (with the DOR notion) is not as effective as we thought.

\begin{table*}[bt!]
\caption{Characterization results for the plane and line examples.}
\label{tab:vcm_1}
\begin{tabular}{ccccccccc}
\hline
& \multicolumn{4}{c}{VCM without DOR notion} & \multicolumn{4}{c}{VCM with DOR notion} \\ 
\cline{2-9} 
& $ColumnSize(J)$ & $Rank(J)$ & DOR & Matched? & $ColumnSize(J)$ & $Rank(J)$ & DOR & Matched? \\ 
\hline
\begin{tabular}[c]{@{}c@{}}The plane example\\ (Representation \#1)\end{tabular} & 16 & 11 & 6 & \XSolidBold & 16 & 11 & 5 & \CheckmarkBold \\
\begin{tabular}[c]{@{}c@{}}The plane example\\ (Representation \#2)\end{tabular} & 24 & 13 & 6 & \XSolidBold & 24 & 13 & 6 & \XSolidBold \\
The line example & 12 & 5 & 6 & \XSolidBold & 12 & 5 & 6 & \XSolidBold \\ 
\hline
\end{tabular}
\end{table*}

The above limitation only represents one limitation found during the authors’ application of VCM. Are there other hidden limitations? At present, the authors do not have a good answer to this question. There are good reasons that it may be better for us to take the negative side: existing work has over-claimed the effectiveness of VCM twice, of which one is before the introduction of DOR, and the other before this work. Despite of these limitations (found ones and future ones), VCM is still a promising (maybe the most promising) method for handling the task of constraint state characterization (other known methods are less effective than VCM and have inherent limitations). Further development of VCM is necessary to prefect VCM (for constraint state characterization). Two improvement directions maybe taken to achieve the goal: (1) have a comprehensive analysis of VCM’s capabilities and limitations, and then device new (add-on) methods to specifically handle the limitations; and (2) propose a variant of VCM for which a rigorous proof is achievable. The authors have presented a systematic method in \parencite{Zou2019variational} to address Limitation \RNum{1}, following the latter direction.

\subsection{Limitation \RNum{2}}

This subsection and the next discuss the limitations of VCM for dealing with the second issue in geometric constraint solving. To repeat, the issue concerns with the optimal decomposition of a non-well-constrained constraint system into the minimal over-constrained parts and the maximal well-constrained parts. The limitation this subsection focuses on is the ineffectiveness of the current methods on minimal over-constrained part detection.

Over-constrained parts in a non-well-constrained GCS could take the form of a group of dependent constraints, or the form of a group of geometric entities whose associated constraints are more than needed. These two forms are equivalent; and this work defines an over-constrained part as a group of dependent constraints. The task of minimal over-constrained part detection is to minimize the size of each over-constrained part in the GCS.

Existing methods to accomplish the task share the same idea of using greedy methods \parencite{thierry2011extensions, michelucci2009interrogating, moinet2014defining, hu2017over}. Conceptually, they first find a maximal subset of independent constraints in the given GCS, then decide the dependencies of the rest constraints on these independent constraints in a one-by-one manner, which can be understood using the simple example in ~Equation (\ref{eq:vcm_1}). To simplify the indexing of these equations, they are numbered $E1$-$E5$ sequentially. The existing methods start from a seed equation, e.g., $E1$, then iterate through all equations to greedily find a maximal subset of independent equations; for this case, it will be $\{E1,E2,E3\}$. Subsequently, the dependencies of the rest constraints with the constraints already presented in the subset will be checked. For example, equation $E4$ is dependent with all the three constraints in the subset, and the same for equation $E5$. According to these dependencies, the methods will output two minimal dependency groups: $\{E1,E2,E3,E4\}$ and $\{E1,E2,E3,E5\}$. 

\begin{equation}
    \begin{cases}
       x+y+z=0,\\
       2x+y+x=1,\\
       3x+2y+z=1,\\
       x+2y+3z=1,\\
       2x+4y+3z=2.
    \end{cases}
    \label{eq:vcm_1}
\end{equation}

The above strategy may fail to detect the minimal over-constrained parts in a GCS because the maximal independent subset of a constraint system is not unique; this non-uniqueness could lead to wrong detection results. Consider the example in ~Equation (\ref{eq:vcm_2}) (a slight modification of ~Equation (\ref{eq:vcm_1})). The existing methods will give two “minimal” dependency groups: $\{E1,E2,E3,E4\}$ and $\{E1,E2,E3,E5\}$. However, it is easy to see that there is another dependency group with smaller size: $\{E4,E5\}$.

\begin{equation}
    \begin{cases}
       x+y+z=0,\\
       2x+y+x=1,\\
       3x+2y+z=1,\\
       x+2y+3z=1,\\
       2x+4y+6z=2.
    \end{cases}
    \label{eq:vcm_2}
\end{equation}

From the above discussion, it is safe to say that the development of current methods for minimal over-constraint detection is far from being sufficient. The insufficiency has its origins in the inherent limitations of greedy methods. 
Again, two directions may be taken to solve the limitation or insufficiency. It can be seen from the examples that greedy methods gave the wrong detection results due to the wrong choice of the seed equation. For example, if the seed equations is set to equations $E4$, the greedy methods may give the right detection results. Therefore, one improvement direction is to have a preprocessing that finds that optimal seed equation. This problem is however a challenging issue for its own sake and remains an open issue. The other improvement direction, which is more desirable, is to put effort into the insights of the detection problem and to attain a solid mathematical modeling of this problem. The authors have presented a new, precise formulation of the minimal over-constraint detection problem in \parencite{Zou2020adecision} to address Limitation \RNum{2}, following the latter direction.

\subsection{Limitation \RNum{3}}

The limitation this subsection discusses is the ineffectiveness of the current methods on maximal well-constrained part detection. Maximizing sizes of well-constrained parts is equivalent to minimizing under-constrained parts because under-constraint is described by DOFs between well-constrained parts. A well-constrained part of a GCS refers to a subset of the participating geometric entities whose induced constraint subsystem is well-constrained. An induced subsystem of a part is the subset of the constraints that are defined merely over the part. A well-constrained part P is deemed as maximal if there is no another subset $P'$ of the participating geometric entities such that $P\subset P'$ and $P'$ is well-constrained.

Existing methods to solve the problem of maximal well-constrained part detection also use greedy methods \parencite{thierry2011extensions, michelucci2009interrogating}, which leads to the limitation similar to Limitation II. The work flow of these methods is as follows. First, a geometric entity set S is initialized with a seed geometric entity (randomly chosen); then, iterate through all the other geometric entities to check if the geometric entity form a well-constrained part with the geometric entities in S, and, if so, add the geometric entity to S; the resulting S gives a maximal well-constrained part. Repeat above procedures on the left geometric entities until no geometric entity is left. Similar to Limitation II, the initialization of the methods — the choice of seed geometric entities — has a significant effect on their effectiveness, and they may fail to detect the maximal well-constrained parts in a GCS.

\begin{figure}[bt!] 
\centering
\includegraphics[width=0.5\textwidth]{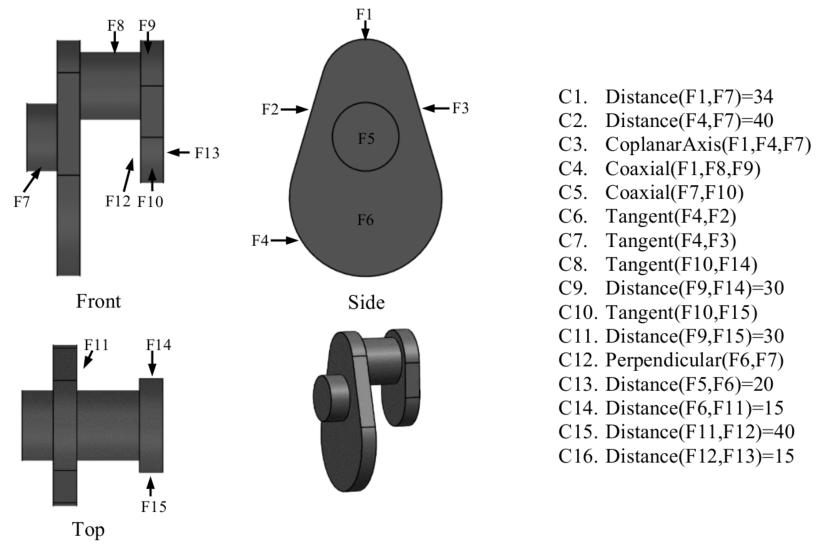}
\caption{An under-constrained GCS example.}
\label{fig:vcm_3}
\end{figure}

\begin{figure}[bt!] 
\centering
\includegraphics[width=0.5\textwidth]{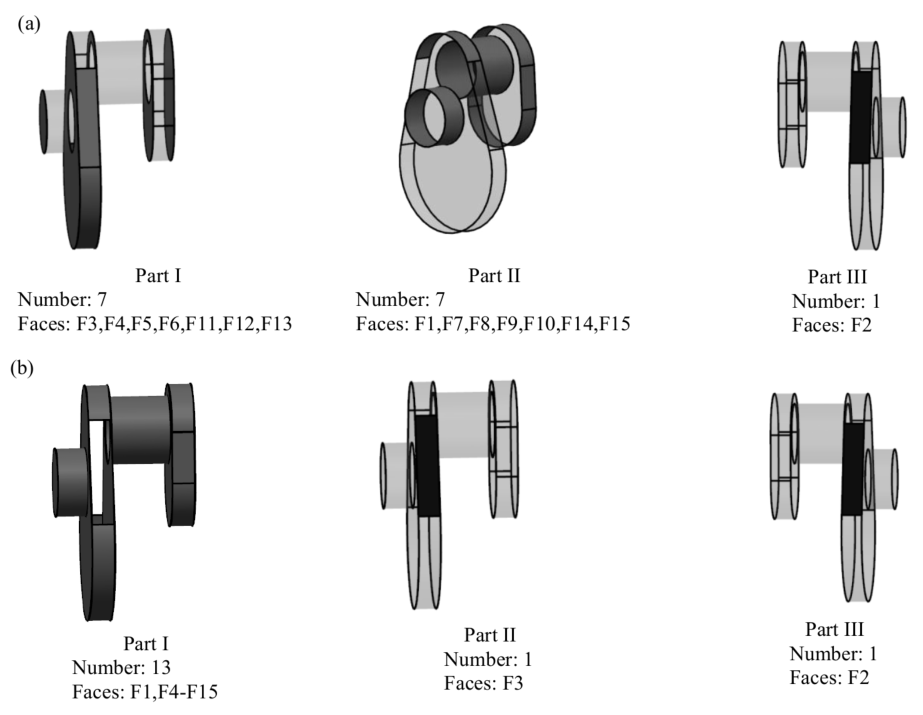}
\caption{Results of maximal well-constrained part detection.}
\label{fig:vcm_4}
\end{figure}

An example of this limitation is given in ~Figure \ref{fig:vcm_3} and ~Figure \ref{fig:vcm_4}. A crank model with two DOFs is shown in ~Figure \ref{fig:vcm_3}, and the two DOFs are a rotation of plane $F2$ about cylinder $F4$’s axis and a rotation of plane $F3$ about cylinder $F4$’s axis. ~Figure \ref{fig:vcm_4}a shows one detection result of using the greedy method, where the seed geometric entity for part \RNum{1} is plane $F3$, the seed geometric entity for part \RNum{2} is cylinder $F7$, and that for part \RNum{3} is plane $F2$. However, ~Figure \ref{fig:vcm_4}b shows another solution to the problem of maximal well-constrained part detection, with a larger size on the well-constrained part \RNum{1}.

This limitation, again, has its origins in the inherent limitations of greedy methods. In fact, the use of greedy algorithms does not give a formulation of the detection problem but represents an incomplete technical tool. To address the issue, we need to develop a mathematical modeling of the maximal requirement on sizes of well-constrained parts. The authors have presented a new, precise formulation of the maximal well-constraint detection problem in \parencite{Zou2020adecision} to address Limitation \RNum{3}.

\subsection{Limitation \RNum{4}}

This limitation is related to the third issue in geometric constraint solving — the optimal decomposition of a well-constrained GCS into the minimal well-constrained parts. It is not strictly necessary to have a VCM-based method to decompose a well-constrained GCS into the minimal well-constrained parts. The major advantage of VCM over the previous graph-based methods \parencite{hoffman2001decomposition1} is the ability to handle constraint dependency, but the setting of the third issue excludes such dependencies as the given GCS is well-constrained. There have been effective methods to accomplish the task of decomposing a well-constrained GCS into the minimal well-constrained parts, such as the graph-based methods \parencite{hoffman2001decomposition1}.

For VCM to be self-contained and systematic, we want to have a VCM-based method to decompose a well-constrained GCS into the minimal well-constrained parts (although this may not have a practical value). To our knowledge, there is no existing work on this topic. This issue, as already noted, is secondary.

\section{Conclusions}

Limitations of VCM have been discussed in this work, under the context of geometric constraint solving. To have a systematic discussion of these limitations, we first outlined the three essential tasks/issues in geometric constraint solving, then introduced the basic idea of VCM in a brief fashion, and finally showed the capabilities and limitations of VCM using examples. The history of geometric constraint solving has also been briefly reviewed. With the limitations stated above, we expect that VCM can be perfected, starting from solving these limitations.

\part{Acknowledgements}

This work is a byproduct of the authors developing a new CAD modeling approach: variational direct modeling \parencite{zou2022variational,qiang2019variational}. The associated project was in part funded by the Natural Sciences and Engineering Research Council of Canada (NSERC). This financial support is greatly appreciated.

\part{Conflict of interest statement}
The authors declared no potential conflicts of interest with respect
to the research, authorship, and/or publication of this article.

\printbibliography

\end{document}